# Reentrant glass transition in a colloid-polymer mixture with depletion attractions

T. Eckert and E. Bartsch

Institut für Physikalische Chemie, Universität Mainz, Jakob-Welder-Weg 15,

D-55099 Mainz, Germany

Performing light scattering experiments we show that introducing short-ranged attraction to a colloidal suspension of nearly hard spheres by addition of free polymer produces new glass transition phenomena. We observe a dramatic acceleration of the density fluctuations amounting to the melting of a colloidal glass. Increasing the strength of the attractions the system freezes into another nonergodic state sharing some qualitative features with gel states occurring at lower colloid packing fractions. This reentrant glass transition is in qualitative agreement with recent theoretical predictions.

Colloidal suspensions have become veritable model systems for the study of fundamental problems of condensed matter science like crystal nucleation [1,2] and the glass transition [3-7]. Especially the possibility to closely approximate hard sphere like behavior with sterically stabilized colloids [3,4] has lead – in combination with specific predictions of the so called mode coupling theory (MCT) for the dynamical behavior of such particles at an ideal glass transition [8] – to a profound improvement of the understanding of the physical processes involved in the glassy freezing of matter [9,10]. So far the particle interactions of the colloidal systems studied at the colloid glass transition were of the purely repulsive type. In all cases the glass physics follows the same scenario: a slowing down of particle dynamics due to the so called cage effect – the trapping of particles by transient cages of neighboring particles - which becomes more efficient with increasing particle density until structural relaxation eventually freezes. This is manifested by a characteristic two step decay of particle fluctuations which has been found for colloids with both short-ranged hard sphere-like [[3,4] and long-ranged Yukawa-like [6] repulsive interactions. The interaction range influences only details like numerical values of critical exponents and glass transition volume fractions.

In contrast, indications exist that short-ranged attractive interactions generate new physics at the glass transition. In a study of the phase behavior of a colloid-polymer mixture where short-ranged attractive interactions between colloidal PMMA spheres are introduced by free non-adsorbing polymer via the well-known depletion effect [11,12] a shift of the glass transition line to higher volume fractions on increasing the strength of the attractions was observed [13]. The glass line seems to meet the gel line which at high attraction strength extends from low to high colloid volume fractions. This raised the so far unsolved question of the relationship between gel and glass states, a problem which has been addressed by solving mode coupling equations for hard sphere systems with short ranged attractions [14-16]. Theory predicts that with increasing strength of the short ranged attractive interactions (i.e. on decreasing the temperature in an atomic system which corresponds to an increase of the polymer concentration in case of depletion attractions) a competition arises between packing effects and bonding effects as the main mechanism of glassy freezing. The bonding effects tend to stabilize the liquid state and the glass transition line recedes to higher packing fractions. As bonding effects start to dominate the packing effects the transition line bends back to lower packing fractions and meets the extension of the gel line. This gives rise to a new phenomenon – the reentrant melting of a glass: there exists a region in the liquid state where glassy freezing occurs on both decreasing and increasing the temperature. The confirmation of this prediction was reported in two molecular dynamics simulation studies for model systems [17,18].

In this letter we report the first observation of this phenomenon for a colloidal suspension by measurements of the full time dependence of the density autocorrelation functions. We present dynamic light scattering data on a colloid-polymer mixture which show that switching on short-ranged depletion attractions by adding free polymer to a colloidal glass dramatically speeds up the density fluctuations and the glass melts. This correlates with a decrease of the structure factor peak and its shift to higher scattering vectors. On further increasing the polymer concentration the dynamics slows down again and the system freezes into another nonergodic state, giving rise to a reentrant glass transition.

The system under study is a binary mixture of polystyrene micronetwork spheres (the polymer chains within the particles are highly crosslinked by one crosslink per 50 monomers) swollen in the good solvent 2-ethyl-naphthalene to radii of $R_{small}$ = 150 nm and $R_{large}$ = 180 nm, respectively. The particles could be reasonably well be approximated as hard spheres which was inferred from the phase behavior, rheology studies, and the static structure factor of the individual components [19]. The total colloid volume fraction, φ, was obtained by setting $\varphi_f = 0.49$ for each of the components at their crystallization concentration and calculating $\varphi$ according to the number ratio $N_{small}/N_{large} = 2.65$ which was kept constant in all experiments. The resulting size polydispersity was about 12 %, sufficient to completely suppress crystallization. In order to introduce short-ranged attractive interactions we added linear polystyrene chains with a narrow molecular weight distribution ($R_g = 8.6$ nm, $M_w = 136000$ g mol$^{-1}$, $M_w/M_n =1.04$). The resulting size ratio $\delta = R_{g,polymer}/\langle R_{colloid}\rangle$ (where $\langle R_{colloid}\rangle$ is the average sphere radius in the binary colloid mixture) was 0.054. The polymer concentrations are given with respect to the free volume available to the polymer chains in terms of the so called reservoir volume fractions $\varphi^R$ [13]. Static structure factors S(q) and density autocorrelation functions f(q,$\tau$) were determined using light scattering equipment and procedures as described previously [5].

The dynamics of the pure binary colloidal mixture closely follows the scenario predicted by the ideal mode coupling theory (MCT) for hard spheres and experimentally found for hard sphere like PMMA colloids (Ref.[4] and literature cited there) as is shown by the thick lines in Fig.1. On increasing the packing fraction beyond $\varphi = 0.55$ the density autocorrelation function develops a two step decay indicative of the well-known cage effect which within the idealized MCT is the essential mechanisms of glassy freezing [8]. On crossing the glass transition line long scale structural rearrangements are arrested and only small scale fluctuations of particles about their metastable equilibrium positions persist. As a consequence, f(q,$\tau$) does not longer decay to zero but to a finite plateau value. This is seen in the change from the curve for $\varphi = 0.592$ to $\varphi = 0.62$, i.e. beyond the glass transition at $\varphi_g = 0.595$ [20]. The remnant decay of the density fluctuations seen at very long decay times $\tau$ for the glassy samples can be attributed to aging effects and has been observed in other systems as well [21]. On further increasing the particle density the amplitude of the local particle movements decreases and the plateau height of f(q,τ) increases until all dynamics comes to a halt at the packing fraction of random close packing $\varphi_{rcp}$. This situation is almost reached for the binary mixture at φ= 0.67. The increase of the glass transition value observed for our system ($\varphi_g$ = 0.595) relative to the one observed for the nearly monodisperse system ($\varphi_g$ = 0.575 [4]) is mainly caused by the higher polydispersity for the system studied here. Similarly, polydispersity shifts $\varphi_{rcp}$ to a higher value as compared to the monodisperse case (where $\varphi_{rcp}$ = 0.64).

If one adds to these (nearly) glassy samples a small amount of linear polystyrene (δ = 0.054) amounting to (reservoir) volume fractions of about 0.255±0.05 one observes a tremendous acceleration of the particle dynamics (thin lines in Fig.1) which corresponds to the melting of the colloidal glass. As increasing the polymer concentration corresponds to a reduction of the temperature in an analogous atomic glass, this phenomenon corresponds for φ = 0.62 and 0.67 to a melting of a glass on cooling. These dramatic changes in the particle dynamics are a direct consequence of subtle changes in the short range order which are induced by the introduction of short ranged attractions. Fig.2 demonstrates these changes in S(q) for the binary hard sphere colloidal mixture with a packing fraction of 0.592. One sees a shift of the peak of the static structure factor S(q) to higher scattering vectors q (signaling a shift of average particles distances to smaller values) and a decrease of the peak height accompanied by a broadening on increasing the polymer concentration. These effects result from an effective attractive force induced between the particles by an unbalance of the osmotic pressure on the colloids due to the depletion of polymer chains between two approaching spheres. This effective "depletion" attraction leads to a larger probability of finding two particles at close contact than in the reference system without free polymer and the peak of S(q) shifts to higher wave vectors. At the same time the higher average number of particles at close contact creates additional volume for other particles thereby increasing their mobility. The additional free volume gives rise to more structural disorder, the height of the main peak of S(q) decreases and it broadens.

The decrease of the peak height, which occurs for glassy samples in the same manner as shown in Fig.2, allows understanding the melting of the colloidal glass within the frame of the MCT. Here, it is the cage effect, i.e. the trapping of particles in transient cages of their next neighbors, which - becoming more effective on increasing the particle density - is at the origin of glassy freezing. The more effective caging correlates with an increase of the peak height of S(q) as it is the density mode corresponding to the maximum of S(q) which is the critical mode driving the slowing down of the density fluctuations. The switching on of depletion attraction leads to pairs (or with less probability triplets, quadruplets, etc.) of neighboring particles being much closer to each other than without attractions, thus "opening gates" in the cages allowing the trapped particles to escape. This increased particle mobility tends to destabilize the glassy state and is at the origin of the "melting of a glass" observed in Fig.1. This mechanism and its effect on short range order and dynamics have been suggested by theory [14,16] and the results shown in Fig.2 are in qualitative agreement with corresponding predictions (see Fig. 2 of [16]). The accompanying decrease of the S(q) peak height thus reflects the fact that to compensate for this increased free volume, the packing fraction has to be increased in order to reach the glass transition line.

One fascinating phenomenon predicted for a system of hard spheres with very short-ranged attractions close to the glass transition is the occurrence of reentrant melting, i.e. the possibility to drive the system into the glass by both cooling and heating. In terms of a colloid-polymer mixture it means that one should observe a slowing down and, eventually, the freezing of colloid dynamics on both increasing and decreasing the polymer concentration. Starting from the hard sphere reference system in the glassy state addition of free polymer should first speed up the dynamics, such that one observes a melting, and should then slow down the particle motions again on approaching the extension of the gel line to higher packing fractions, resulting in the transition into another nonergodic state. That this behavior can indeed be found in our system is shown in Fig.3, where the corresponding density autocorrelation functions $f(q,\tau)$ and structural relaxation times $\tau_\alpha$ (approximately determined by $f(q,\tau_\alpha) = 0.3$) are depicted for the binary mixture with a colloid packing fraction of 0.67. Starting well in the glass state at $\varphi^R = 0$ (see upper curve in Fig.1 ), addition of free polymer to yield $\varphi^R = 0.11$ (Fig.3a, curve 1) softens the glass as indicated by the reduction of the plateau value (from about 0.96 to 0.9). A further increase of attractions leads to glass melting and produces the characteristic two step relaxation pattern of the "supercooled" liquid ($\varphi^R = 0.25$, curve 2). At $\varphi^R = 0.47$ (curve 3) the system is in the liquid state. Adding more polymer ($\varphi^R = 0.55$, curve 4) the colloid dynamics slows down again and between $\varphi^R = 0.6$ and 0.76 (curves 5 and 6) the transition line into another nonergodic state is crossed. Clearly visible in Fig.3 is the qualitative difference between the glassy state at high polymer concentration (where bond formation due to depletion attraction is important) and the nearly hard sphere glass at low polymer concentration (caused by hard sphere repulsion effects), if one compares the glassy samples with $\varphi^R = 0.11$ and $\varphi^R = 0.76$ as well as the "supercooled" liquids at $\varphi^R = 0.21$ and $\varphi^R = 0.6$, respectively. Moving from $\varphi^R = 0.21$ in direction of decreasing polymer concentration one sees the appearance of the two step decay of $f(q,\tau)$ characteristic of the cage effect, a feature which is still visible deep in the glass state. Compared to this the density autocorrelation curves remain rather structureless, showing almost monotonous decays, on approaching the polymer-rich glass.

This difference, which expresses the signatures of the two different glassy states of matter, which may be called "packing-driven" and "bonding-driven" glass, has been predicted by theory [16] and can be qualitatively understood by the presence or absence of the cage effect. Loosely speaking, the plateau value of f(q,τ) is a measure of the probability that caged particles in a "packing-driven" glass undertake excursions over a distance 2π/q around their (metastable) equilibrium positions, the probability increasing with decreasing plateau value. This probability is significantly reduced in the "bonding-driven" glass when probing the dynamics on the length scale $2\pi/q_{max}$ (corresponding to the cages of the "packing-driven" glass) due to strong bonding effects which tend to keep particles fixed at contact with their next neighbors. The corresponding plateau value moves close to 1 and becomes undetectable within experimental accuracy. In other words, increasing the tendency to form more bonds between particles by increasing the particle attractions imparts to the "packing-driven" glass some feature characteristic of ordinary gels, namely the randomness of bond formation with the concomitant loss of the short range order connected with the presence of transient cages, and a new type of amorphous state is formed: the "bonding-driven" glass. This disturbance of the "packing-driven" glass structure is limited at high colloid volume fractions due to the presence of excluded volume effects. On reducing the strength of excluded volume effects by going to lower colloid volume fractions the transition line of the "bonding-driven" glass seems to smoothly change into the gel line. At least this is suggested by the experimental phase behavior [13] and the theoretical results of MCT [16] as well as its extensions into the gel region [15].

We have shown by adding increasing amounts of polymer chains to a hard sphere colloid glass that the introduction of short ranged depletion forces result in reentrant melting followed by the formation of an "bonding-driven" glass state. The scenario observed is in very good agreement with predictions of MCT for hard spheres with short ranged attraction forces.

The authors thank W. Götze, M. Sperl and M. Fuchs for stimulating discussions, A. M. Puertas et al. for providing a preprint of their simulation work prior to publication and H. Sillescu for critical comments. Support of the Deutsche Forschungsgemeinschaft (DFG) within the Sonderforschungsbereich (SFB) 262 is gratefully appreciated.

**Figure captions**

Fig.1 Comparison of the density autocorrelation functions $f(q,\tau)$ of a hard sphere colloidal suspension (see text) before (thick solid lines) and after (thin solid lines) the addition of linear polymer chains (size ratio $\delta = R_{g,polymer}/\langle R_{colloid}\rangle = 0.054$). The colloid volume fractions $\varphi_{colloid} = 0.592$, 0.62 and 0.67 of each set of $f(q,\tau)$ increase from left to right as indicated in the figure. The dynamics is probed at a scattering vector corresponding to the peak maximum of $S(q)$ of the pure colloid suspension at its glass transition volume fraction $\varphi_g \approx 0.595$ [20]. One clearly sees a dramatic speeding up of the structural relaxation on addition of free polymer (reservoir volume fractions $\varphi^R_{polymer} = 0.26$) which amounts to the melting of a colloidal glass upon switching on the depletion attraction.

Fig.2 Changes induced in the static structure factor $S(q)$ of a colloid suspension (see text) closely below the glass transition ($\varphi_{colloid} = 0.592$) by switching on depletion attractions and increasing the interaction strength by adding various amounts of free polystyrene chains (volume fractions $\varphi^R_{polymer} = 0, 0.21, 0.47$ as indicated).

Fig.3 Effect of increasing the free polymer volume fraction $\varphi^R_{polymer}$ on the dynamics of a colloid suspension in a glassy state close to random close packing ($\varphi_{colloid} = 0.67$) indicating the phenomenon of reentrant melting.
  a) density autocorrelation functions $f(q,\tau)$ (probed at the same scattering vector as in Fig.1). The polymer reservoir volume fractions $\varphi^R$ are indicated by the numbers at the curves: (1) 0.11, (2) 0.21, (3) 0.47, (4) 0.55, (5) 0.59, (6) 0.76. The thick solid lines correspond to nonergodic (glassy) samples [20].

b) structural relaxation times $\tau_\alpha$ as determined by $f(q,\tau_\alpha) = 0.3$ versus polymer reservoir volume fraction $\varphi^R_{polymer}$. The vertical lines schematically indicate the transition lines to the respective nonergodic states. Note that samples with $\varphi^R_{polymer}$ = 0.12 and 0.76 appear nonergodic within the time window of the light scattering experiment [20], even though the correlation functions seem to indicate a decay at long times. The latter can be attributed to aging effects.

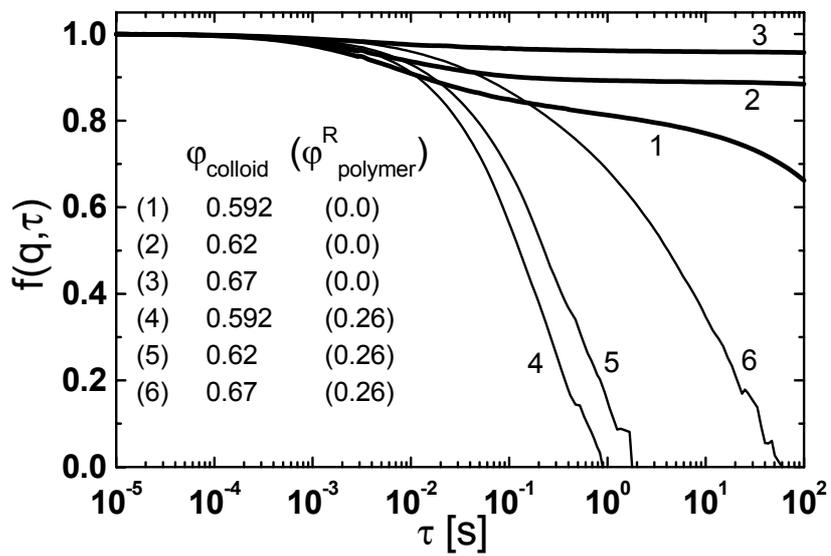

**Fig.1**

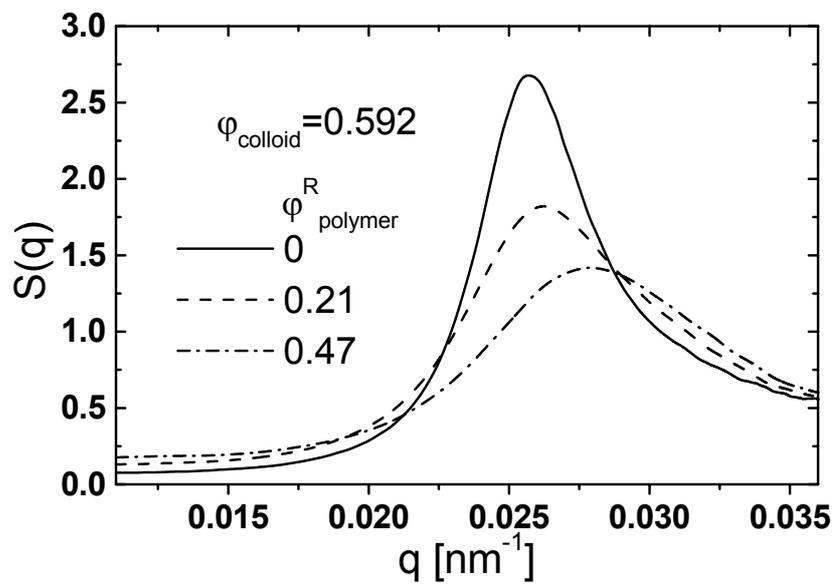

**Fig.2**

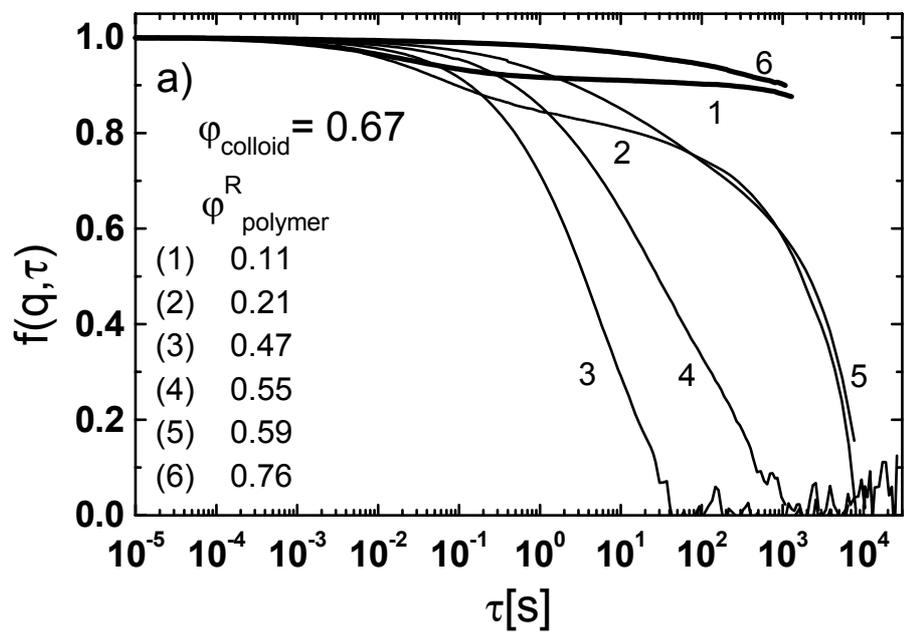

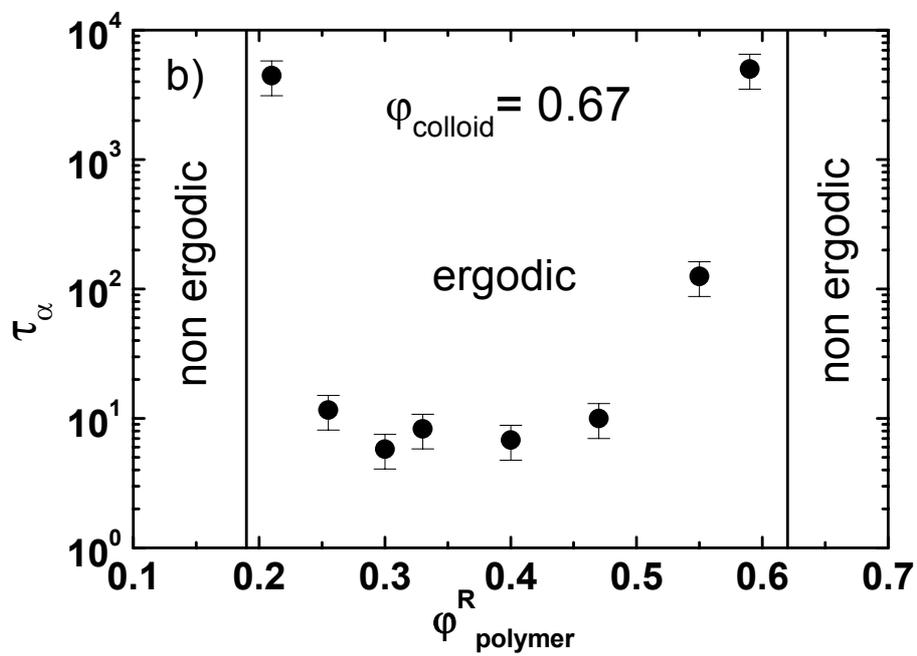

**Fig.3**